\documentclass[onecolumn]{revtex4}
\usepackage[dvips]{graphicx}

\begin{document}

\title{Constraints on Primordial Black Holes by Distortions of Cosmic
  Microwave Background}

\author{Hiroyuki Tashiro}
\email{hiroyuki.tashiro@ias.u-psud.fr}
\affiliation{
Institut d'Astrophysique Spatiale, Universit\'e Paris-Sud XI, 
B\^atiment 121, Orsay, F-91405, France}
\author{Naoshi Sugiyama}
\email{naoshi@a.phys.nagoya-u.ac.jp}
\affiliation{Department of Physics and Astrophysics, Nagoya
  University, Chikusa-ku, Nagoya,  464-8602, Japan}
\affiliation{Institute for Physics and Mathematics of the Universe, University of Tokyo, 5-1-5 Kashiwa-no-Ha, Kashiwa City,
Chiba 277-8582, Japan}

\date{\today}

\begin{abstract}
Possible influence of primordial black hole (PBH) evaporations 
on cosmic microwave background (CMB) is investigated.  
The spectrum distortions of 
CMB from the black-body spectrum are described by the chemical potential $\mu$
and the Compton parameter $y$. From COBE/FIRAS limits
on $\mu$ and $y$, the power law index $n$ of primordial
density fluctuations and the mass fraction of PBHs $\beta$ are
constrained by employing the peak theory for the formation process of
PBHs. Constraints set here are $n < 1.304$ and $n<1.333$ in the
thresholds of peaks $\zeta_{\rm th} =0.7$ and $\zeta_{\rm th} =1.2$,
respectively, for the PBH mass range between $2.7\times 10^{11}$g and
$1.6 \times 10^{12}$ g, and $n < 1.312$ and $n<1.343$ in the thresholds
of peaks $\zeta_{\rm th} =0.7$ and $\zeta_{\rm th} =1.2$,
respectively, for the PBH mass range between $1.6 \times 10^{12} ~{\rm
g}$ and $3.5\times 10^{13}$ g, which correspond to the comoving scales
between $3 \times 10^{-18}$ Mpc and $ 4\times 10^{-17}$ Mpc. The
constraint on the PBH fraction, which is the direct probe of the
amplitude of density fluctuations in these scales, stays at almost
the same value as $\beta<10^{-21}$ in these mass ranges. It is also found
that, with these constraints, UV photons injected by PBH
evaporations are unlikely to ionize the majority of hydrogen atoms.

\end{abstract}

\pacs{98.70.Vc ;98.80.Cq }

\maketitle

\section{Introduction}
The properties of the primordial density fluctuations in large scales
are revealed by Wilkinson Microwave Anisotropy Probe (WMAP) and galaxy redshift surveys \cite{wmap,2df,sdss}.
The results of these observations are essentially consistent with the
predictions of the inflationary model, i.e., primordial density
fluctuations with an almost scale-invariant spectrum and the random
Gaussian statistics.

Taking the second order terms of the inflaton potential and density
perturbations into account, however, we expect to have the departure
from the scale invariant spectrum and the existence of the
non-Gaussian components. Therefore, precise measurements of the spectrum
and the non-Gaussian components are crucial for a detailed understanding of
inflation. For example, the first year WMAP data together with
galaxy redshift surveys and Ly-$\alpha$ measurements preferred
the non-zero running spectral index \cite{wmap},
while the third year data with new Ly-$\alpha$
measurements do not strongly favor the running \cite{spergel}. 

The Ly-$\alpha$ measurements are sensitive to the power spectrum at
$\sim 1 ~{\rm Mpc}$, while measurements of strong gravitational lensing
have a sensitivity on much smaller scales such as $\sim 10$-$100 ~{\rm kpc}$.
These methods are, however, limited in both scales and epochs. So far
the best probe of the small scale density fluctuations is provided
by the abundance of primordial black holes (PBHs)~\cite{pbh}.
%
%
The formation of PBHs took place during the radiation dominated epoch
due to the gravitational collapse of the high density region at 
the horizon scale when the amplitude of over-density exceeded a
critical threshold. Therefore the resultant mass spectrum and
the abundance of PBHs depend on the amplitude of the power spectrum for 
the primordial density fluctuations at the horizon crossing epoch.  
It is known that PBHs eventually evaporate while emitting Hawking
radiation~\cite{hawkingrad}. The lifetime of a PBH is proportional
to the cubic of its mass. Therefore, PBHs with mass less than $10^{15}$g should
have evaporated away by the present epoch. When they evaporate, they emit 
black-body radiation and numerous kinds of particles such as neutrinos,
electrons and protons.  

Since PBHs with mass larger than $10^{15}$g survive in the present
universe, we can set a constraint on their abundance from the
fact that PBH density cannot exceed the average matter density
observed at present. We can therefore introduce $\beta(M)$ which is the
fraction of the regions of mass $M$ collapsing into PBHs~\cite{beta}.  
This can be related to the mass fraction of PBHs at the
time of formation as $\beta=\rho_{{\rm PBH},i}/\rho_{r,i}$ where
$\rho_{{\rm PBH},i}$ and $\rho_{r,i}$ are the energy densities of PBHs
and background radiation at the PBH formation epoch, respectively.
The constraint on the density parameter of PBHs at present $\Omega_{{\rm
PBH},0}<1$ implies $\beta < 10^{-18} (M/10^{15}{\rm g})^{1/2}$.

Another important constraint on the abundance of PBHs can be set
by the phenomena of evaporation. PBHs emit many kinds of particles when they
evaporate. Among these are diffuse gamma rays.  
The emitted photons
from evaporated PBHs, after the recombination epoch, contribute to the
diffuse gamma rays. The upper limit on diffuse gamma rays 
sets a stringent constraint on the abundance of PBHs~\cite{gamma}.  
It is also known that big bang
nucleosynthesis~\cite{nucle} and entropy production in the early
universe also provide constraints on PBH abundance for various
mass ranges (see the review \cite{abundance}).

In this paper, we investigate spectrum distortions of 
cosmic microwave background radiation (CMB) caused by PBH evaporation.
Naselskii was the first to
study the effect of PBHs on the recombination process and to estimate
the allowed abundance of PBHs that evaporated around the
recombination epoch~\cite{distort}. Naselskii and Shevelev 
estimated distortions of CMB due to electrons and positrons from PBH evaporations \cite{naselskii-shevelev}.
Ricotti et al. have also investigated the effects of non-evaporating PBHs on CMB
\cite{ricotti}.
X-rays emitted by gas accretion into non-evaporating PBHs were found to modify
recombination and reionization. 
Therefore, by studying the effect of such X-rays on the $y$-distortion and the reionization, 
they obtained a PBH's constraint which is larger 
than $10^{15} M_\odot$. 
Here, we adopt a
modern analysis of CMB spectrum distortions, i.e., $\mu$ and $y$
distortions on which COBE/FIRAS has set stringent observational limits,
where $\mu$ and $y$ are the chemical potential and the Compton $y$-parameter, 
respectively. 
%
The process we consider here is that photons which have evaporated from PBHs
directly induce the $\mu$-distortion in CMB and also hit electrons in
the surrounding medium. Accordingly, these electrons scatter CMB photons
and generate the $y$-distortion. While electrons directly emitted
by PBHs also generate the $y$-distortion via inverse Compton
scattering, resultant photons are too energetic to be observed as CMB
distortions in COBE/FIRAS data. Therefore we ignore electron emission
from PBHs in this paper.

Distortions are created when the energy injection into CMB
arises between redshift $z\sim 10^6$ and recombination $z\sim
1000$. Before $z \sim 10^6$, CMB can recover the black-body distribution
via double Compton scattering or free-free emission. After
recombination ($z < 1000$), there is almost no free electron to be a
target of photons.
Therefore we calculate CMB distortions caused by photons emitted by 
PBHs in this epoch.  
Consequently, we can obtain the constraint on the evaporated PBH
abundance using COBE/FIRAS limits, $|\mu|<9\times 10^{-5}$ and $y<1.5
\times 10^{-5}$ \cite{firas}.  
Note that the mass range of PBHs constrained here 
is between $10^{11}$ and $10^{13}$g, since this mass range corresponds to 
the PBHs whose lifetimes are within $1000<z< 10^6$.


In order to estimate the number density of PBHs, we need the threshold
value of density fluctuations required for the overdense regions to collapse into
PBHs. The threshold value was traditionally calculated as the critical
density contrast in the simplified cosmological model by the analytic
method, where the value obtained is $\delta_c = 1/3$ for the radiation
dominant era \cite{pbh}. There has been resent progress in 
estimating the threshold value by employing numerical simulations 
based on general relativity~\cite{n-j, s-s}.  
In this paper, therefore, we adopt the threshold value of Shibata and Sasaki
\cite{s-s}.  

This paper is organized as follows. In Sec. II we review the formula for
evaluating the PBH mass function by using the peak theory. In Sec. III we
calculate CMB distortions, $\mu$ and $y$, 
caused by Hawking radiation from PBHs  
and we set constraints on the primordial power
spectral index and the PBH abundance.  
Conclusions are presented in Sec. IV. In this paper we assume that
$h=0.70 \ (H_0=h \times 100 ~{\rm km/s \cdot Mpc})$, $\Omega _{\rm b}
h^2 =0.022$ and $\Omega_{\rm M} h^2 =0.11$ \cite{spergel}. And we set
$c=\hbar=1$, where and $c$ and $\hbar$
are the speed of light and Planck's constant over $2\pi$, respectively.

\section{PBH mass function}

In this section, we calculate the distribution of PBHs produced from
the primordial density fluctuations by following the procedure of
Green et al.\cite{g-etal}. We adopt the threshold value of density
fluctuations required for the overdense regions to collapse into PBHs in the
horizon crossing epoch obtained by Shibata and Sasaki~\cite{s-s} in
which they performed numerical simulation of evolution for the metric
perturbations on the uniform-expansion hyper surface. Green et
al. \cite{g-etal} related this threshold to the gauge-invariant
curvature perturbation on the uniform-density hyper surface, $\zeta $,
which is defined as \cite{invariant}
\begin{equation}
\zeta ={\cal R} - H {\delta \rho \over \dot \rho},
\label{zeta-def}
\end{equation}
where $\rho$ is the background density, $\delta \rho$ is the perturbed
density, ${\cal R}$ is the curvature perturbation and dot represents
the time derivative. Shibata and Sasaki's threshold depends on the
environment of the PBH formation. The threshold value $\zeta _{\rm
th}$ is $\zeta _{\rm th}=0.7$ when the density peak is surrounded by a
low density region. In contrast, if the peak is surrounded by a flat
Friedman-Robertson-Walker region, $\zeta _{\rm th}=1.2$.

Green et al. advocated the peak theory for the calculation of the PBH
mass function because Shibata and Sasaki's result is described as the
constraint on the peak value of fluctuations. We assume a
power-law primordial power spectrum, ${\cal P}_{{\cal R}} ={{\cal
R}_c} (k/k_0)^{n-1} $. From WMAP and galaxy survey results, 
${{\cal R}_c}$ is $(24.0 \pm 1.2) \times
 10^{-10}$ at the scale $k_0 =0.002 {\rm Mpc}^{-1}$ \cite
{spergel}. We smooth the density fields by a Gaussian window function
with comoving size $R$. Under these assumptions, the peak theory
gives the comoving number density of the peaks which are higher than
$\nu$ as \cite{g-etal}
\begin{equation}
n (\nu, R) =
{1 \over (2 \pi)^2} {(n-1)^{3/2} \over 6^{3/2} R^3} {(\nu^2 -1)}
\exp\left(-{\nu^2 \over 2}\right),
\label{density-peak}
\end{equation}
where we employ the high peak limit $\nu\gg 1$. 
We relate $\nu$ to the threshold value of the PBH formation by
\begin{equation}
\nu =
\left[{2 (k_0 R )^{n-1} \over {{\cal R}_c} \Gamma ((n-1)/2) }\right]^{1/2}
\zeta _{\rm th}.
\label{nu-threshold}
\end{equation}
We can assume Eq. (\ref{density-peak}) to be the comoving number density
of PBHs which are formed from the overdense regions with scale $R$.

Let us relate the smoothing scale $R$ to the PBH mass. The PBH mass
depends on the initial environment around the peak as well as the
threshold value. For simplicity, however, we assume that PBHs with
the horizon mass are produced when the overdense regions enter the
horizon. The epoch when scale $R$ crosses the horizon is
evaluated as
\begin{equation}
R= {1 \over a H}.
\label{formation-time}
\end{equation}
From the comoving entropy conservation, we obtain the relation between
the temperature $T$ and the scale factor $a$ (which is normalized at
the present epoch as $a_0=1$)
\begin{equation}
g_* ^{1/3} a T= const.,
\label{temperature-scale} 
\end{equation}
where $g_*$ is the number of relativistic degrees of freedom. In the
radiation era, the horizon mass at horizon crossing of the comoving
scale $R$ is described from Eqs. (\ref{formation-time}) and
(\ref{temperature-scale}) as
\begin{eqnarray}
M_{\rm BH} (R) &=&
{4 \pi \over 3} \left({8 \pi G \over 3}\right)^{-1}
\left[ {H_0^2 \Omega_{\rm M} \over 1+z_{\rm eq}} \left({g_{* \rm eq} 
\over g_*}\right)^{1/3}\right]^{1/2} 
R^2
\nonumber \\
&=&
10^{15}  \left( g_* \over 100 \right)^{-1/6}
\left( R\over 6.2 \times 10^8\ {\rm cm}\right)^2 ~{\rm g}. 
\label{bh-mass}
\end{eqnarray}
Here, we assume that the redshift of the matter-radiation equality is
$z_{\rm eq}=3200$ \cite{wmap} and the number of relativistic degrees
of freedom at the equality is $g_{* \rm eq}=3.36$.

The formation epoch of a PBH with the mass $M_{\rm BH}$, 
which we assume as the horizon crossing epoch of this mass scale, 
can be 
described in terms of the scale factor as 
\begin{equation}
a_{\rm BH}(M_{\rm BH}) = \left(g_{*{\rm eq}} \over g_* \right)^{1/6} 
H_0 \Omega_{\rm m} ^{1/2} \left(M_{\rm BH} \over X \right)^{1/2}, 
\end{equation}
\begin{equation}
X={4 \pi \over 3} \left({8 \pi G \over 3}\right)^{-1}
\left[ {H_0^2 \Omega_{\rm M} \over 1+z_{\rm eq}} \left({g_{* \rm eq} \over g_*}\right)^{1/3}\right]^{1/2}.
\end{equation}
Moreover, employing Eq. (\ref{bh-mass}), we can rewrite
Eq. (\ref{density-peak}) as a function of the PBH mass $n(\nu, M)$.

To describe the PBH abundance, one often uses the fraction of regions
of mass $M$ which collapse to PBHs, $\beta(M)$. Obviously, $\beta(M)$
also represents the mass fraction of PBHs at the time of formation.
The energy density of PBHs with mass $M_{\rm BH}$ at 
horizon crossing of scale $R$ is expressed as $\rho_{\rm BH}(M_{\rm
BH})= M_{\rm BH} n(\nu, M_{\rm BH}) a_{\rm BH}( M_{\rm BH})^3$. 
The mass fraction of PBHs $\beta$ can be written as 
\begin{equation}
\beta(M_{\rm BH})={\rho_{\rm BH}(M_{\rm BH}) \over \rho}
={1 \over (2 \pi)^2} {(n-1)^{3/2} \over 6^{3/2}} {(\nu^2 -1)}
\exp\left(-{\nu^2 \over 2}\right).
\label{pbh-beta}
\end{equation}
Here we adopt 
the high peak approximation, Eq. (\ref{density-peak}).

One could conclude that the total number density of PBHs with mass $M_{\rm
BH}$ at a given epoch can be described as $n (\nu, M_{\rm BH})$.
However, this is the number density of PBHs when they were formed.
For the total number at a given epoch, which we express as $n_{\rm
BH} (\nu, M_{\rm BH}, z)$, we need to subtract the number of PBHs which have been
absorbed by larger PBHs in subsequent epochs. To take into account this effect, 
we consider the continuity equation for the PBH number density. 
The comoving 
number density of PBHs at $z$, in the range from $M_{\rm BH}$ to $M_{\rm
BH} + dM_{\rm BH}$ can be described, in the high peak limit, as
\begin{eqnarray}
n_{\rm BH} (\nu,M_{\rm BH}) dM_{\rm BH}&=& -{d n(\nu, M_{\rm BH}) \over dM_{\rm BH}} dM_{\rm BH}
\nonumber\\
&=& {1 \over 4 \pi ^2 M_{\rm BH}}  \left(X (n-1) \over 6 M_{\rm BH} \right)^{3/2}\nonumber\\   
&&~~~~\times \left[
 {(n-1) \over 2 }\nu^4 (n,M_{\rm BH} )- {3 \over 2} (n-3) \nu^2 (n,M_{\rm BH})-3
\right]
\exp \left(-{\nu^2 (n,M_{\rm BH}) \over 2}\right) 
\nonumber\\
&\approx&
{1 \over 4 \pi ^2 M_{\rm BH}}  \left(X (n-1) \over 6 M_{\rm BH} \right)^{3/2}   
{(n-1) \over 2 }\nu^4 (n,M _{\rm BH}) \exp \left(-{\nu^2 (n,M_{\rm BH}) \over 2}\right),
\label{pbh-numberdensity}
\end{eqnarray}
\begin{equation}
\nu (n,M_{\rm BH}) = \left[ 2 \left(k_0 ^2 M_{\rm BH} / X  \right)^{(n-1)/2} 
\over {\cal R}_c \Gamma \left((n-1)/2 \right) \right]^{1/2} \zeta _{\rm th} .
\end{equation}
At the third step in Eq. (\ref{pbh-numberdensity}), we take the high
peak limit, $\nu \gg 1$.

Let us discuss the epoch of first PBH formation. For
simplicity, we set the end of inflation at this epoch. Accordingly, 
the initial condition of PBH formation can be characterized by 
the reheating temperature. Under this assumption, 
the minimum mass of PBHs depends on 
this temperature. The comoving horizon
scale at this epoch is expressed from
Eqs. (\ref{formation-time}) and (\ref{temperature-scale}) as
\begin{eqnarray}
{1\over aH}&=&\left[H_0^2 \Omega_{\rm M} (1+z_{\rm eq}) 
\left({g_* \over g_{* \rm eq}}\right)^{1/3 } \right]^{1/2}
\left( T_{\rm eq} \over T_{\rm rh} \right) ,
\nonumber \\
&=&2.0 \times 10^8 \left({g_* \over 100}\right)^{-1/6}  \left(T_{\rm rh}\over 10^8 {\rm GeV} \right)^{-1}
~{\rm cm},
\label{horizon-temperature}
\end{eqnarray}
where $T_{\rm rh}$ is the reheating temperature. By using
Eq. (\ref{bh-mass}), we obtain the minimum mass of PBHs as
\begin{equation}
M_{\rm BH,min}(T_{\rm rh}) = 9.8\times 10^{15}  {\left(T_{\rm rh}\over 10^8 {\rm GeV} \right)^{-2} }~{\rm g}.
\label{eq:BH_min}
\end{equation}
In our assumption, therefore, $n(\nu,M)=0$ if $M<M_{\rm BH,min}(T_{\rm rh})$.

\section{Constraints from the CMB distortions}

PBHs inject energy into CMB through Hawking radiation. In the
early universe, even when there is a large amount of energy
injection, CMB achieves black-body spectrum by
photon-electron interaction, i.e., Compton and double Compton
scatterings \cite{lightman}. However, decouplings of these interactions occur one by
one after $z \sim 10^6$ so that the distortions from the black-body are
produced via the energy injection.

First, the double Compton scattering is decoupled at $z \sim 10^6$.
After this decoupling, the total photon number between CMB and the
injected photons conserves. Accordingly, CMB can no longer achieve
the black-body spectrum if there are any photon injections after the
decoupling of the double Compton scattering, although the injected
photons are still thermalized by the Compton scattering. The CMB
spectrum in this thermal equilibrium state is described as the
Bose-Einstein spectrum with the chemical potential $\mu$ which stands
for ``distortions from the black-body spectrum''.

Following the double Compton decoupling, the decoupling of the
thermalization comes at $z \sim 10^5$. Due to the expansion of
the universe, Compton scattering does not work effectively and can
no longer establish the thermalization when the time scale of the
thermalization becomes longer than the Hubble time. We must describe
the distortion from the black-body spectrum due to the photon
injections as the Compton-$y$ parameter once the decoupling of the
thermalization takes place.

These distortion parameters are constrained by COBE/FIRAS,
$|\mu|<9\times 10^{-5}$ and $y<1.5 \times 10^{-5}$.
In this section
we discussion constraints on the PBH abundance or the spectral index 
of density fluctuations 
by the limits of the chemical potential and the Compton-$y$ parameter.

\subsection{$\mu$-distortion}

When energy $Q$ is injected continuously into CMB,
time evolution of the chemical potential, 
which describes the distortion of the CMB energy spectrum, 
is given as \cite{h-s}
\begin{equation}
{d \mu \over dt} = - {\mu \over t_{\rm DC} (z)} +1.4 {Q \over \rho_\gamma}.
\label{mu-evolution}
\end{equation}
Here $t_{\rm DC}$ is the time scale for the double Compton scattering
\begin{equation}
t_{\rm DC} = 2.06 \times 10^{33} (1-Y_{\rm p}/2)^{-1}(\Omega_{\rm b} h^2)^{-1} z^{-9/2}
~{\rm s},
\label{double-compton-time-scale}
\end{equation}
where $Y_{\rm p}$ is the primordial helium mass fraction.
The solution of Eq. (\ref{mu-evolution}) is given by Hu and Silk \cite{h-s} as
\begin{equation}
\mu = 1.4 \int^{t(z_{\rm freeze})} _0 dt 
{Q \over \rho_\gamma} \exp \left[- \left(z \over z_{\rm DC}\right)\right],
\label{mu-value}
\end{equation}
\begin{equation}
z_{\rm DC} = 1.97 \times 10^{6} \left( 1-{1 \over 2}\left({Y_{\rm p} \over 0.24} \right) \right)^{-2/5}
\left( \Omega_{\rm b} h^2 \over 0.0224 \right)^{-2/5},
\label{z-mu}
\end{equation}
\begin{equation}
z_{\rm freeze} = 2.86 \times 10^{5} \left( 1-{1 \over 2}\left({Y_{\rm p} \over 0.24} \right)\right)^{-1/2}
\left( \Omega_{\rm b} h^2 \over 0.0224 \right)^{-1/2},
\label{z-free}
\end{equation}
where $z_{\rm DC}$ is the characteristic redshift for decoupling of
the double Compton scattering and $z_{\rm freeze}$ is the redshift
when the injected energy can no longer be thermalized.

As mentioned before, the $\mu$-distortion takes place if there are any
photon injections during the epoch after decoupling of the double
Compton scattering and before decoupling of the Compton scattering, i.e., 
$z_{\rm freeze} <z < z_{\rm DC}$.


Let us now estimate the energy injection $Q$ due to 
the Hawking evaporation of PBHs.
A Schwarzschild black hole with mass $M$
emits particles with spin $s$ and total energy between $E$ and $E+dE$ 
at a rate per degree of freedom 
\begin{equation}
{dN_{\rm emit} \over dt dE} dE={\Gamma_s  \over 2 \pi \hbar}
\left[ \exp \left( E \over kT(M) \right) -(-1)^{2s} \right]^{-1} dE,
\label{emit-rate}
\end{equation}
where $T(M)$ is the temperature of a black hole with mass $M$ 
and written as
\begin{eqnarray}
T(M) &=&{1 \over 8 \pi G  M} \nonumber\\ 
&\approx& 1.0 \left({M \over 10^{13} \rm g} \right)^{-1} ~{\rm GeV} .
\label{temperature-mass} 
\end{eqnarray}
Here $\Gamma_s$ is the dimensionless absorption probability of emitted species.
For simplicity, we assume that only photons contribute to $Q$.  
In the case of photons, $\Gamma_s$ is described as \cite{mac-w}
\begin{equation}
\Gamma _s = \left\{
\begin{array}{ll}
{64 G^4 M^4 E^4 /3  },  & E\ll kT(m), \\
{27 G^2 M^2 E^2 },   &E\gg kT(m).
\end{array}
\right.
\end{equation}

The PBH mass $M$ decreases due to the Hawking evaporation.  
The mass loss rate of a PBH can be written as \cite{macgibbon}
\begin{equation}
 {d M \over dt} = -5.34 \times 10^{25} f(M) M^{-2} {\rm g \ sec}^{-1},
\label{mass-loss-rate} 
\end{equation}
where $f(M)$ is a function of the number of species which are directly
emitted, and can be described by the fitting formula as
\begin{eqnarray}
f(M )&=&1.569+0.569 
\left[ \exp\left({-0.0234 \over T(M)}\right) +6 \exp\left({-0.066 \over T(M)}\right)
+3 \exp\left({-0.11 \over T(M)}\right)+ \exp\left({-0.394 \over T(M)}\right) \right.
\nonumber \\
&& \qquad
\left. +3 \exp\left({-0.413 \over T(M)}\right)+3 \exp\left({-1.17 \over T(M)}\right)
+3  \exp\left({-22 \over T(M)}\right) \right] +0.963  \exp\left({-0.10 \over T(M)}\right).
\end{eqnarray}
Here $T(M)$ is written in the unit of $\rm GeV$ as in 
Eq.~(\ref{temperature-mass}).

Since $f(M)$ is a weak function of $M$, we
can approximately integrate Eq. (\ref{mass-loss-rate}) and obtain 
the time evolution of the PBH mass with an initial mass
$M_{\rm BH}$ by
\begin{equation}
M(M_{\rm BH}, t) \approx \left[ M_{\rm BH} ^3- 1.5\times 10^{26} f(M)t \right] ^{1/3},
\end{equation}
where the typical values of $f(M)$ are 1.0, 1.6, 9.8 or 13.6 for $M\gg
10^{17}$g, $M=10^{15}$g, $\ 10^{13}$g or $10^{11}$g, respectively. From
Eq. (\ref{mass-loss-rate}), we can also obtain the life time $\tau$ of
the PBH with the initial mass $M_{\rm BH}$ ,
\begin{equation}
\tau (M_{\rm BH}) = 1.87 \times 10^{-27} ~\int^{M_{\rm BH}} dM {M^2 \over f(M)} ~{\rm s }.
\label{lifetime}
\end{equation}

Finally we can write the energy injection rate due to PBHs
as
\begin{equation}
{\dot Q(t)}=\int^{M_H(t)} _{M_{\rm min}(t)} dM_{\rm BH} 
\int^{\infty}_{0} dE  a^{-3} (t)  n_{\rm BH} (M_{\rm BH})
 {dN_{\rm emit} \over dt dE}( M(M_{\rm BH},t)) E,   
\label{energy-injection}
\end{equation}
where $M_H (t)$ is the horizon mass at $t$
and $M_{\rm min}(t)$ is the minimum initial mass 
of the PBHs with a lifetime $\tau=t$.
The $M_{\rm min}(t)$ can be evaluated by taking the inversion of 
Eq. (\ref{lifetime}).

Now we are ready to calculate the evolution of the chemical potential
induced by photon injections from PBHs. Let us first assume that the
lifetime of the smallest PBHs which are formed right after the
inflation epoch is shorter than the time scale of the double Compton
decoupling $t_{\rm DC}$. Under this assumption, we expect the
existence of PBHs that evaporated away at the epoch of the double
Compton decoupling. The mass of these PBHs is obtained as $M_{\rm DC}
\equiv 2.7\times 10^{11}$g, which corresponds to the mass scale of the
horizon when the temperature $T_{\rm DC} \equiv 3.1 \times
10^{11}$GeV. If the reheating temperature of inflation is higher
than this temperature, i.e., $T_{\rm rh} \gg T_{\rm DC}$, the mass of
the smallest PBHs formed after inflation is less than $M_{\rm DC}$.

\begin{figure}[tb]
\begin{minipage}{.45 \linewidth}
  \begin{center}
    \includegraphics[keepaspectratio=true,height=45mm]{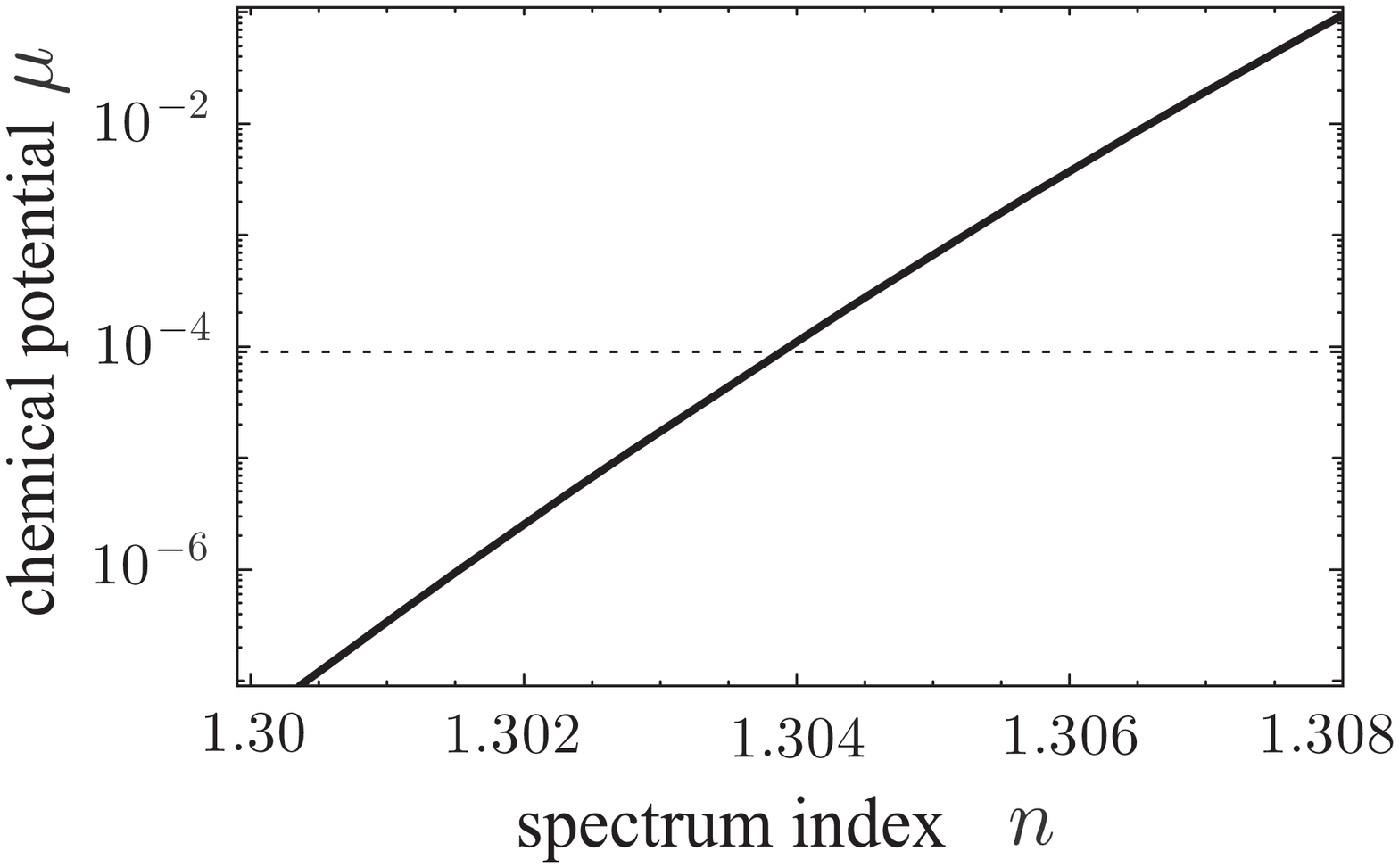}
  \end{center}
\end{minipage}  
\begin{minipage}{.45 \linewidth}
  \begin{center}
    \includegraphics[keepaspectratio=true,height=45mm]{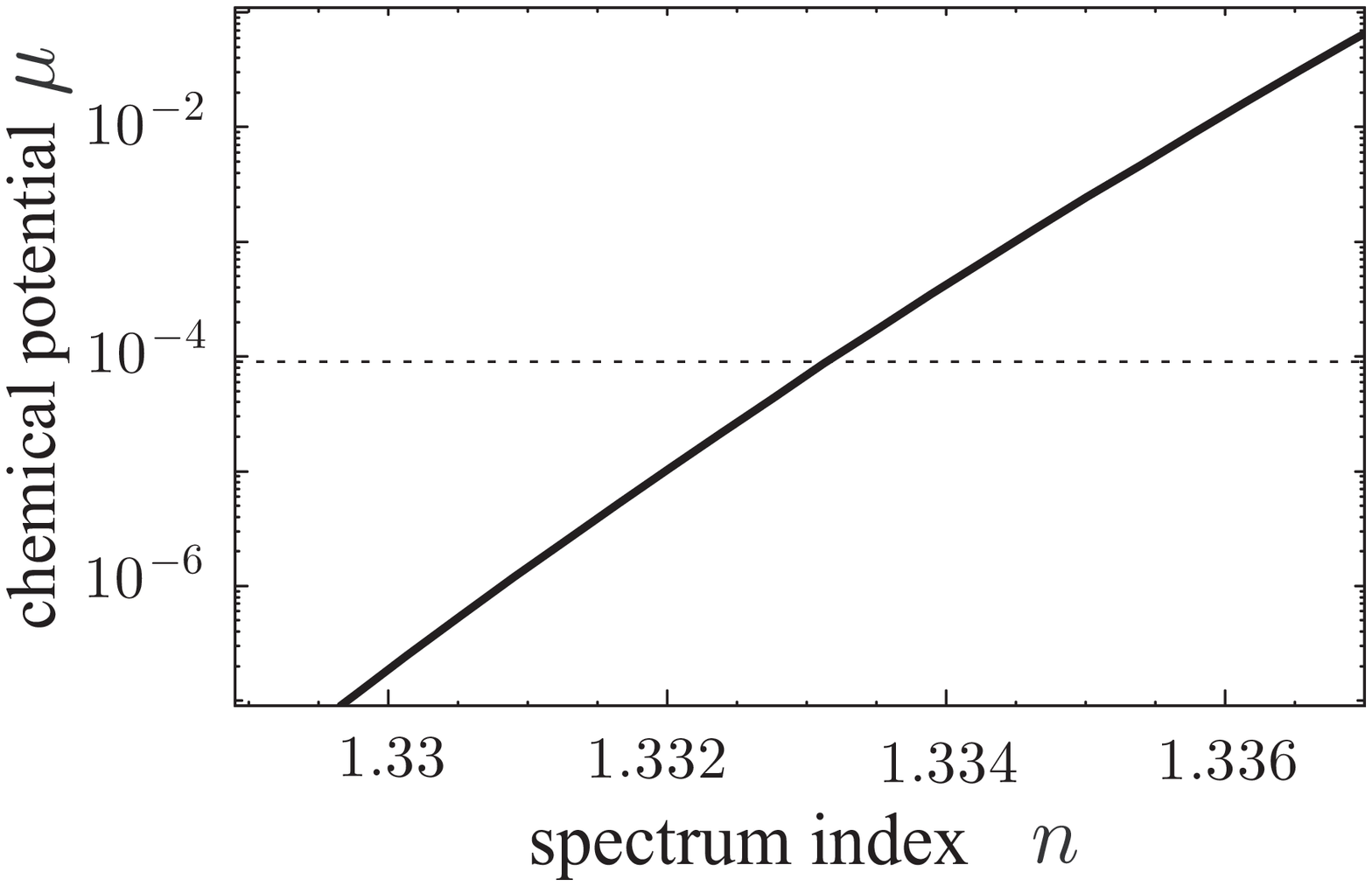}
  \end{center}
\end{minipage} 
  \caption{The chemical potential for each spectral index. We assume
  the reheating temperature $T_{\rm rh} \gg T_{\rm DC}$ (see text).
  The critical thresholds of PBH formation are taken as $\zeta_{\rm
  th} =0.7$ and $1.2$ in the left and right panels, respectively. The
  left and right panels correspond to the cases of PBH formation at
  the density peaks surrounded by the average and low density
  regions, respectively. The dotted line is the upper limit of the
  COBE/FIRAS observation, $\mu =9.0 \times 10^{-5}$. The region
  under the dotted line is allowed. We find the upper limit on the
  spectral index $n$ as $n<1.304$ for $\zeta_{\rm th} =0.7$ and as
  $n<1.333$ for $\zeta_{\rm th} =1.2$.}
  \label{fig:mu.eps}
\end{figure}

Under this assumption, we can compute the chemical potential $\mu$ by using
Eqs. (\ref{mu-evolution}) and (\ref{energy-injection}). We show the
resultant chemical potential as a function of the primordial spectral
index of the power spectrum in FIG. \ref{fig:mu.eps}. The left and
right panels correspond to the cases of the critical threshold of PBH
formation $\zeta_{\rm th} =0.7$ and $\zeta_{\rm th} =1.2$. The
COBE/FIRAS upper limit on the chemical potential is $\mu<9.0 \times
10^{-5}$. We acquire the constraints on the spectral index $n$
as $n<1.304$ for $\zeta_{\rm th} =0.7$ and as $n<1.333$ for $\zeta_{\rm
th} =1.2$.

\begin{figure}[htbp]
  \begin{center}
    \includegraphics[keepaspectratio=true,height=50mm]{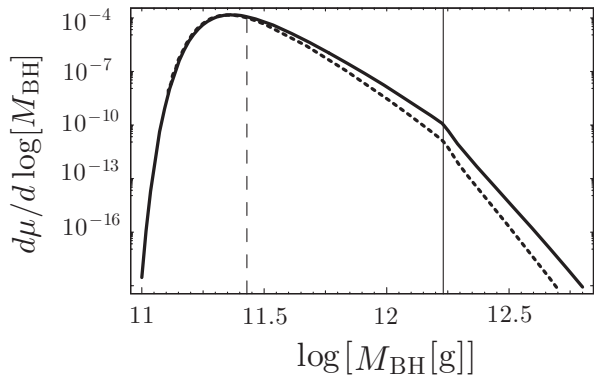}
  \end{center}
  \caption{The production rate of the chemical potential per PBH mass.
  We assume the reheating temperature $T_{\rm rh} \gg T_{\rm DC}$ (see
  text).  The thick solid and thick dotted lines correspond to 
  the cases with $\zeta_{\rm th} =0.7$ and $n=1.304$, and 
  $\zeta_{\rm th} =1.2$ and $n=1.333$, respectively.  
  The vertical thin dashed line represents the characteristic PBH mass scale
  $M_{\rm DC}$ while the thin solid line represents $M_{\rm freeze}$.  
  It is clear that 
  PBHs with the mass around $M_{\rm DC}$ give a dominant
  contribution.}
    \label{fig:mupermass.eps}
\end{figure}

Fig. \ref{fig:mupermass.eps} shows the production rate of the chemical
potential per PBH mass. The thick solid
line is the chemical potential per PBH mass for $\zeta_{\rm th} =0.7$
and $n=1.304$ and the thick dotted line is for $\zeta_{\rm
th} =1.2$ and $n=1.333$. The vertical thin dashed line is for $M_{\rm
DC}$.  
We can see that the majority of contributions to the chemical potential is made by
PBHs with mass around $M_{\rm DC}$.  
This may look counterintuitive because
Eq. (\ref{mu-evolution})
implies that, the larger the PBH mass is, the larger the chemical
potential per evaporated PBH is produced.  
However the number density of PBHs with a smaller mass is much
larger than that of PBHs with a larger mass.  
From the comparison of these two contributions, we find that PBHs 
with the smallest mass existing at $z_{\rm DC}$, which is $M_{\rm DC}$, 
provide most of the contributions for the power spectrum with $n > 1$. 
Accordingly, PBHs with masses below $M_{\rm DC}$, which evaporated away
before $z_{\rm DC}$, are not relevant for the $\mu$-distortion as is shown 
in Fig. \ref{fig:mupermass.eps}.  
Since the reheating temperature only determines the minimum mass of PBHs, 
the constraint on the spectral index from the $\mu$-distortion does not 
depend on the value of the reheating temperature as far as $T_{\rm rh} 
\gg T_{\rm DC}$.  


The other characteristic PBH mass scale is $M_{\rm freeze}\equiv 1.6
\times 10^{12} \rm g$, which corresponds to the mass scale of the
horizon when the temperature $T_{\rm freeze} \equiv 7.6 \times 10^{10}$GeV.  
PBHs with $M_{\rm freeze}$ evaporate away at
the epoch of the thermalization decoupling $z_{\rm freeze}$. This
mass scale is represented as a vertical thin solid line in
Fig. \ref{fig:mupermass.eps}. We can see that the contribution of
PBHs on the chemical potential $\mu$ is quickly suppressed once the
mass exceeds $M_{\rm freeze}$. This feature, which is 
represented by the kink in
each thick line in Fig. \ref{fig:mupermass.eps} at $M_{\rm freeze}$,
can be easily explained by the fact that PBHs with mass below $M_{\rm freeze}$
can only partially evaporate by $z_{\rm freeze}$, and contribute little 
to the $\mu$-distortion.

\begin{figure}[htbp]
  \begin{center}
    \includegraphics[keepaspectratio=true,height=50mm]{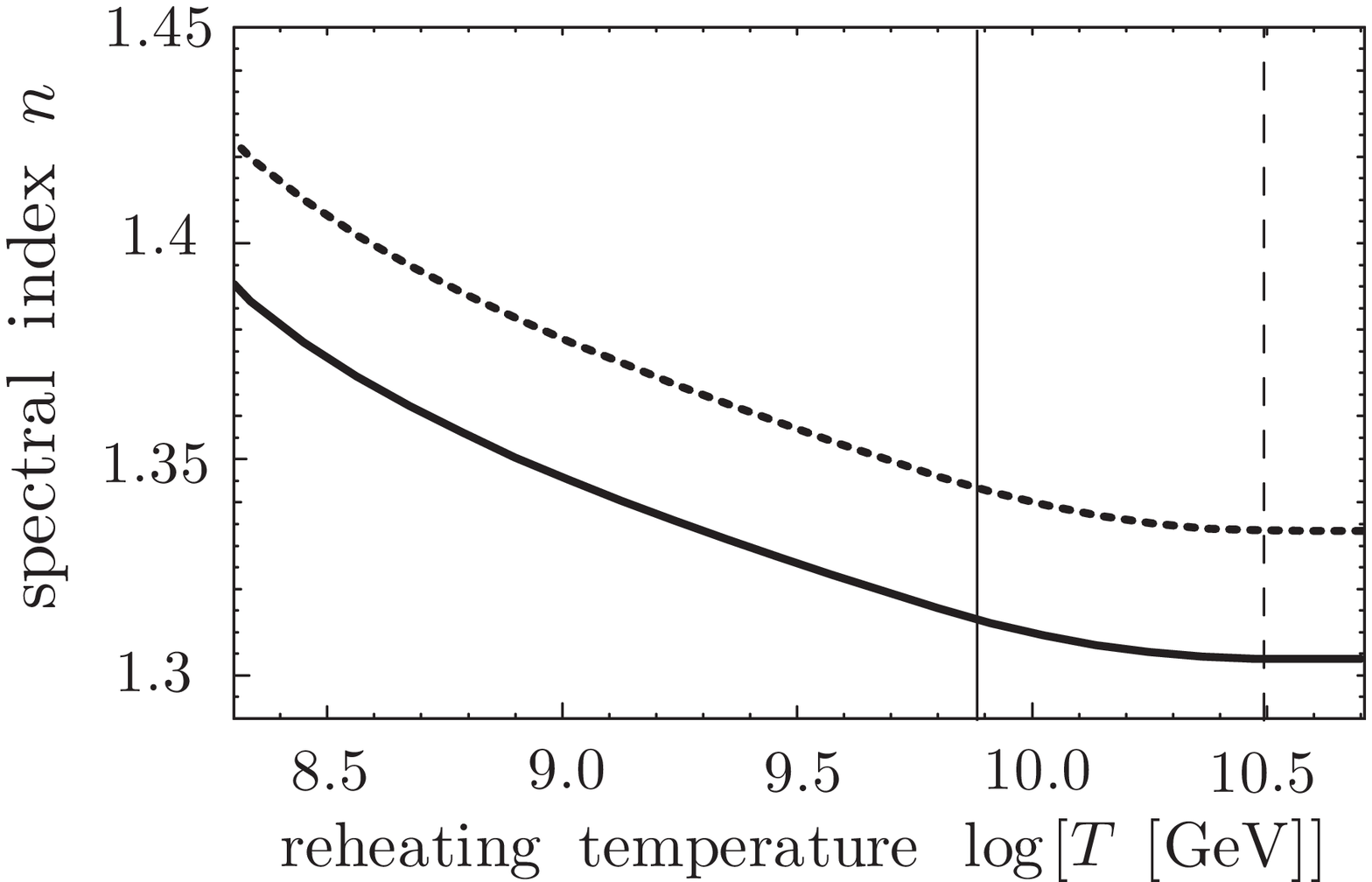}
  \end{center}
  \caption{The constraint on the spectral index as a function of the
   reheating temperature from the $\mu$-distortion.  The thick
   solid line and the thick dotted line are the upper limits with the
   PBH critical threshold $\zeta_{\rm th} =0.7$ and $\zeta_{\rm
   th} =1.2$, respectively.  The vertical thin dashed and thin solid lines 
   represent the reheating temperatures whose horizon scales 
   correspond to PBHs with masses $M_{\rm DC}$ and $M_{\rm freeze}$, 
   respectively. }
  \label{fig:mureheat.eps}
\end{figure}

Let us next consider the case of the reheating temperature $T_{\rm
rh} \lesssim T_{\rm DC}$. In this case, the mass of the smallest PBHs
formed after inflation, which is described as $M_{\rm BH,min}$, is
greater than $M_{\rm DC}$. Accordingly, there is almost no evaporation
of PBHs at $z_{\rm DC}$. 
Instead,
the evaporation takes place at a later epoch. 
Generally speaking, the lower the reheating temperature is, the
later the PBH evaporation takes place. Therefore, we expect that the
constraint on the power law index $n$ becomes looser for a lower
reheating temperature. This behavior is shown in
Fig. \ref{fig:mureheat.eps}.  
Note that if the reheating temperature is so small that
$M_{\rm BH,min}(T_{\rm rh}) > M_{\rm freeze}$, the majority of PBHs 
evaporate away after $z_{\rm freeze}$. Accordingly, the
constraint on the power law index becomes much looser as is shown in
the left hand side of the vertical solid line of
Fig. \ref{fig:mureheat.eps}.

\begin{figure}[htbp]
  \begin{center}
    \includegraphics[keepaspectratio=true,height=50mm]{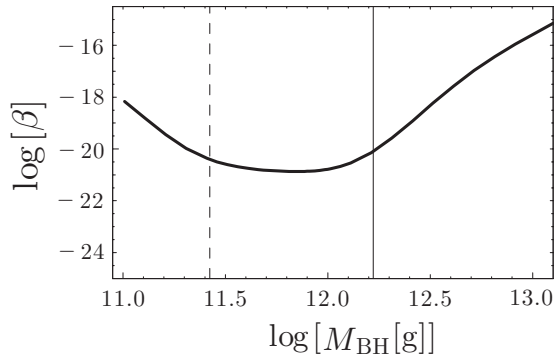}
  \end{center}
  \caption{The constraint on the PBH abundance from the $\mu$-distortion.
  The allowed region is under the thick line. 
  The vertical thin dashed 
  and thin solid lines represent the locations of $M_{\rm DC}$ and
  $M_{\rm freeze}$, respectively.
  }
  \label{fig:mubeta.eps}
\end{figure}

Now, we will constrain the mass fraction of PBHs $\beta$. If $T_{\rm rh}
\lesssim T_{\rm DC}$, we find that greatest contribution to $\mu$
is made by PBHs with the mass $M_{\rm BH,min}$. In this case,
therefore, the constraint on the spectral index for a given reheating
temperature shown in Fig.~\ref{fig:mureheat.eps} can be converted
into the mass fraction of the PBHs with the minimum mass. Describing the reheating
temperature by $M_{\rm BH,min}$ with Eq. (\ref{eq:BH_min}), we obtain
the constraint as shown in Fig.~\ref{fig:mubeta.eps}. On the other
hand, if $T_{\rm rh} \gg T_{\rm DC}$, the most efficient PBHs for the
$\mu$-distortion are those with a mass of $M_{\rm DC}$. Therefore
the constraint from $\mu$-distortion does not provide sufficient clues 
for understanding the
mass fraction on the scale below $M_{\rm DC}$. In
Fig.~\ref{fig:mureheat.eps}, however, we extrapolate our procedure to
slightly smaller masses, since contributions from 
masses slightly smaller than $M_{\rm DC}$ may still be important as is shown in
Fig.~\ref{fig:mupermass.eps}.
The constraint we obtain is: $\beta <
10^{-21}$ between $M_{\rm DC}<M< M_{\rm freeze}$, which 
is tighter than previous constraints found in 
\cite{abundance}.


\subsection{Compton $y$-parameter}

After $z_{\rm freeze}$, 
the injected
photons are no longer thermalized by the Compton scattering. These
photons scatter electrons and the resultant energetic electrons
scatter the CMB photons through the inverse Compton process.
Accordingly, Compton $y$-distortions are induced.

When $T_e \gg T$, 
where $T_e$ and $T$ are electron and CMB temperatures, respectively,  
the Compton $y$-parameter is described from the Kompaneets equation as \cite{s-z}
\begin{equation}
y=\int_{t(z_{\rm freeze})}^{t_0} dt {kT_e \over m_e } n_e \sigma_T ,
\label{y-para-def}
\end{equation}
where $n_e$ and $\sigma_T$ are the electron number density and the
Thomson scattering cross section, and $t(z_{\rm freeze})$ and $t_0$ are
the freeze out time and the present time, respectively.

The electron temperature $T_e$ is controlled 
by Compton heating and Compton cooling as 
\begin{equation}
3 {d \over dt} kT_e n_e =n_e \sigma_T {E_\gamma -4 k T_e \over m_e }
E_\gamma n_\gamma -4n_e\sigma_T k(T_e-T){\pi^2 \over 15} \left({kT
}\right)^3 \left( {kT \over m_e }\right) -15{\dot a \over a} k T_e
n_e,
\label{electron-temperature}
\end{equation}
\begin{equation}
{\partial n_\gamma \over \partial t} = 
{E_\gamma -4 k T_e \over m_e } E_\gamma {\partial n_\gamma \over \partial E_\gamma} n_e \sigma_T
+{2 E_\gamma -4 k T_e \over m_e } E_\gamma  n_\gamma n_e \sigma_T 
+ {\dot a \over a} E_\gamma {\partial n_\gamma \over \partial E_\gamma}
-2 {\dot a \over a}  n_\gamma
+ \delta n_\gamma (E_\gamma),
\label{photon-number-eq}
\end{equation}
where $n_\gamma (E_\gamma)$ is the injected photon number density with
the energy being from $E_\gamma$ to $E_\gamma+dE_\gamma$, and $\delta
n_\gamma (E_\gamma)$ is the production rate of injected photons.  The
first term of the right hand side (rhs) of
Eq. (\ref{electron-temperature}) describes the Compton heating of
electrons by injected photons, the second term describes the Compton cooling of
electrons from CMB photons and the third term describes the adiabatic cooling
due to cosmic expansion.  The first two terms of the rhs of
Eq. (\ref{photon-number-eq}) describe the Compton cooling for injected
photons, and the third and forth terms describe the adiabatic cooling.

For obtaining the electron temperature $T_e$, we employ the analytic
approach in Ref. \cite{f-k}.  When we assume a high electron
temperature ($T_e \gg T$) and the steady state ($dT_e /dt =0$), 
we obtain the electron temperature from Eq. (\ref{electron-temperature})
as
\begin{equation}
kT_e ={m_e  \over 4 \rho_\gamma} 
\int ^{\infty} _{0} {E_\gamma -4 k T_e \over m_e } E_\gamma n_\gamma. 
\label{electron-temp-1}
\end{equation}
Here we ignore cosmic expansion.

Integrating Eq. (\ref{photon-number-eq}) over the photon energy,
we can obtain the equation for the total photon energy density,
\begin{equation}
{\partial \over \partial t} \int^{\infty}_{0} dE_\gamma E_\gamma n_\gamma 
=-\int^{\infty}_{0} dE_\gamma {E_\gamma -4 k T_e \over m_e } E_\gamma n_\gamma n_e \sigma_T 
+{\dot Q},
\label{total-photon-energyrate}
\end{equation}
where we ignore cosmic expansion again because the time scale of the
Thomson scattering is shorter than the cosmological time.  The second
term ${\dot Q} = \int dE_\gamma E_\gamma \left(\delta n_\gamma(E_\gamma)/dE_\gamma \right)$ 
on rhs is the
total injected energy rate and is described as Eq. (\ref{energy-injection})
because we consider the PBHs as the only source of the energy
injection.  The first term on the rhs of
Eq. (\ref{total-photon-energyrate}) describes the energy loss rate per
unit time and, hereafter, we express this term as $E_{\rm loss}$.
From Eq. (\ref{electron-temp-1}), we can rewrite the electron
temperature in terms of $E_{\rm loss}$ as
\begin{equation}
k T_e = {m_e  \over 4 \rho_\gamma}{E_{\rm loss} \over n_e \sigma_T } .
\label{electron-temp-2}
\end{equation}

In order to acquire the electron temperature, we need to estimate the
energy loss rate $E_{\rm loss}$.  A photon with energy $E_\gamma$
loses the energy $E_\gamma (E_\gamma -4kT_e )/m_e $ per Compton
scattering.  Hence, the energy loss of a photon 
with high initial energy $E_{\gamma0} \gg kT_e$ 
within Hubble time is approximated as
\begin{equation}
\delta E_\gamma \simeq {E_{\gamma0}^2 \over m_e} {n_e \sigma_T  \over H}.
\end{equation}

Substituting the Hubble time and the typical photon energy from 
PBHs, our present target of interested, we obtain $\delta
E_{\gamma} \gg E_{\gamma0}$.  This implies that the injected photon
energy is fully transferred into electrons in the Hubble time.  
Therefore we can approximate the energy loss rate $E_{\rm loss}$ as $E_{\rm
loss} \simeq {\dot Q}$.  Substituting ${\dot Q}$  for $E_{\rm loss}$ in
Eq. (\ref{electron-temp-2}), and Eq. (\ref{electron-temp-2}) for
Eq. (\ref{y-para-def}), we obtain
\begin{equation}
y=\int^{t(z_{\rm rec})}_{t(z_{\rm freeze})} dt {{\dot Q} \over 4 \rho_r},
\label{y-dis}
\end{equation}
where the upper bound of the integration $t(z_{\rm rec})$ is the
recombination epoch, which is introduced since the injected energy
does not transfer into the background electrons once the optical depth
becomes very low after recombination.

\begin{figure}[tb]
\begin{minipage}{.45 \linewidth}
  \begin{center}
    \includegraphics[keepaspectratio=true,height=45mm]{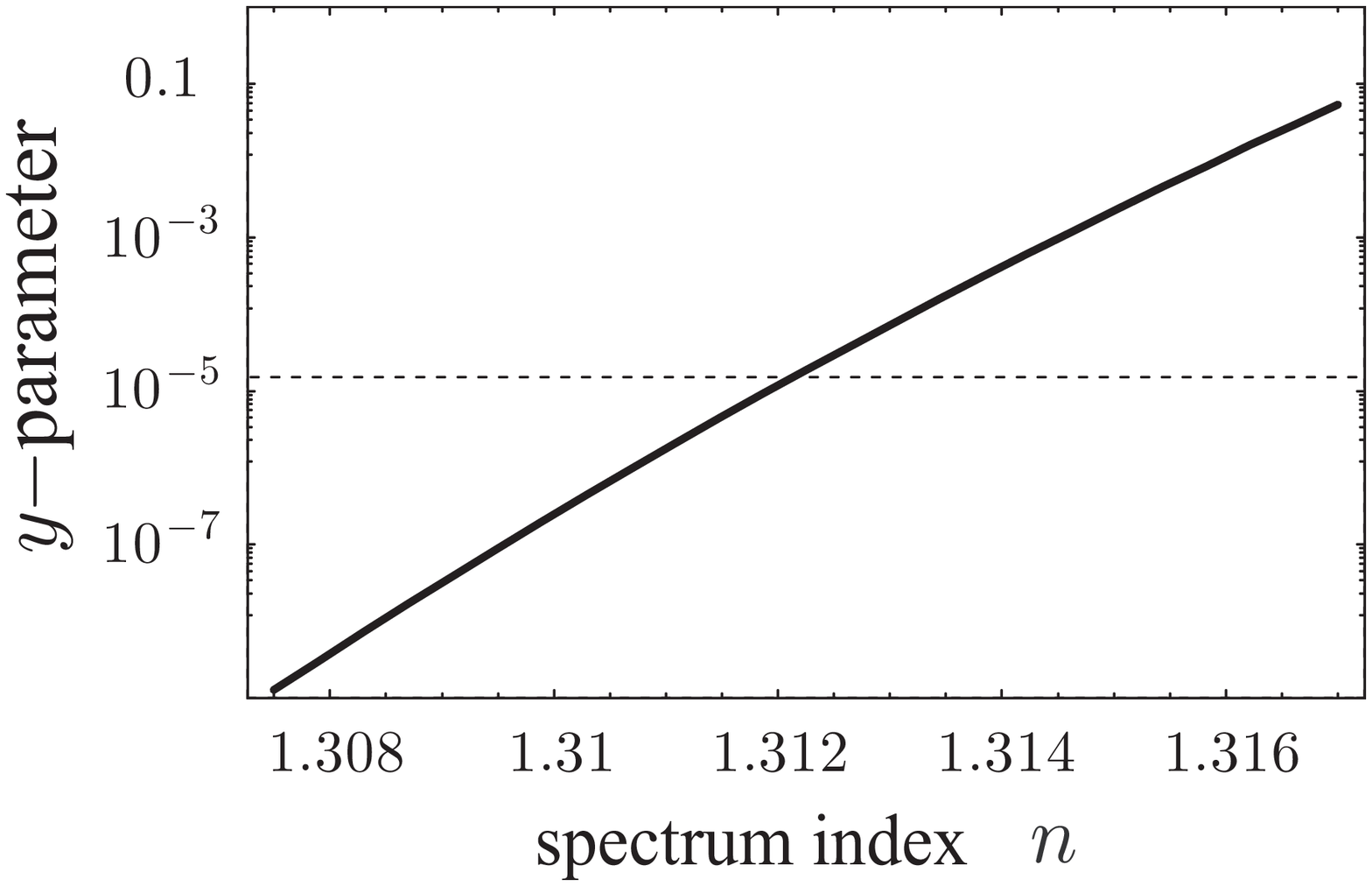}
  \end{center}
\end{minipage}  
\begin{minipage}{.45 \linewidth}
  \begin{center}
    \includegraphics[keepaspectratio=true,height=45mm]{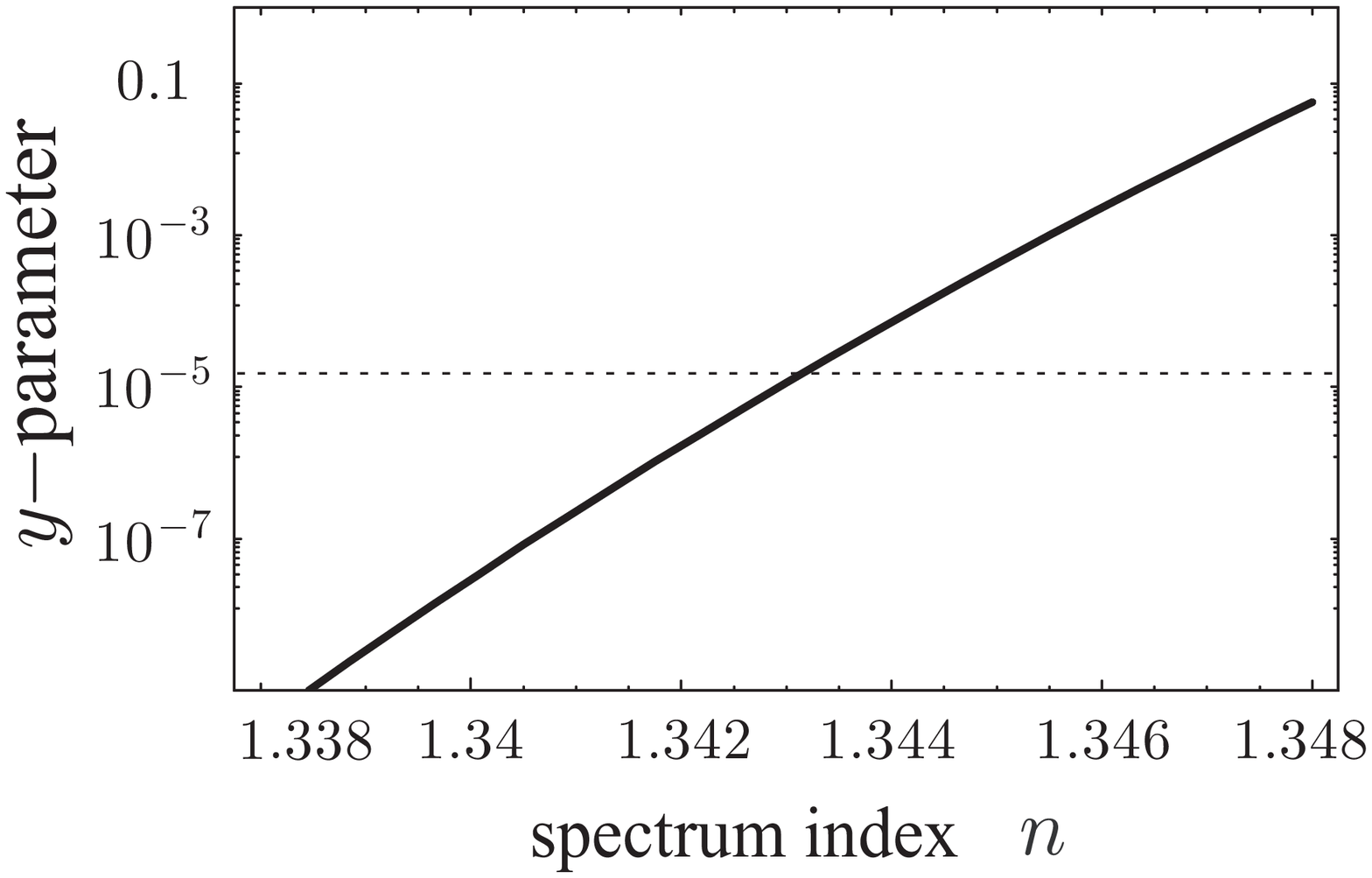}
  \end{center}
\end{minipage} 
  \caption{The $y$-distortion as a function of the spectral index. We
  assume $T_{\rm rh} \gg T_{\rm freeze}$.  The critical thresholds of
  PBH formation are taken as $\zeta_{\rm th} =0.7$ and $1.2$ in the
  left and right panels, respectively. The dotted line is the upper
  limit of the COBE/FIRAS observation. The region under the dotted
  line is allowed. The upper limits on the spectral index $n$ are
  $n<1.312$ for $\zeta_{\rm th} =0.7$ and $n<1.343$ for $\zeta_{\rm
  th} =1.2$.}
  \label{fig:yy.eps}
\end{figure}

\begin{figure}[htbp]
  \begin{center}
    \includegraphics[keepaspectratio=true,height=50mm]{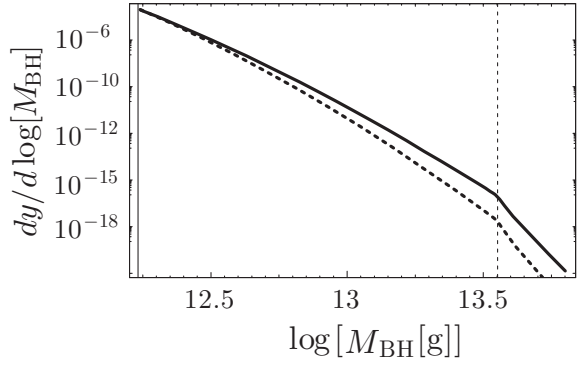}
  \end{center}
  \caption{The production rate of the $y$-parameter per PBH mass. We
  assume $T_{\rm rh} \gg T_{\rm freeze}$. The thick solid and thick
  dotted lines correspond to the cases with $\zeta_{\rm th} =0.7$ and
  $n=1.312$, and $\zeta_{\rm th} =1.2$ and $n=1.343$, respectively.
  The vertical thin solid and dotted lines represent $M_{\rm freeze}$
  and $M_{\rm RC}$, respectively. It is shown that the contribution
  of the PBHs with mass $M_{\rm freeze}$ gives a dominant contribution
  so that the abundance of the PBHs with the mass $M_{\rm freeze}$. }
   \label{fig:yypermass.eps}
\end{figure}

We calculate the $y$-distortion for each primordial power spectral
index under the assumption that the reheating temperature is much
higher than $T_{\rm freeze} \equiv 7.6 \times 10^{10}$GeV which
corresponds to $M_{\rm freeze}$.  The result is shown in
Fig. \ref{fig:yy.eps}.  The upper limit on the $y$-parameter obtained by
COBE/FIRAS is $y=1.5 \times 10^{-5}$ which is the dotted line in
Fig. \ref{fig:yy.eps}.  Therefore, we obtain the constraint on the
spectral index $n<1.312$ for $\zeta_{\rm th} = 0.7$ or
$n<1.343$ for $\zeta_{\rm th} = 1.2$.  The $y$-parameter
per PBH mass is plotted in Fig. \ref{fig:yypermass.eps}.  The thick
solid line is the $y$-parameter per PBH mass with 
$\zeta_{\rm th} =0.7$ and $n=1.312$, and the thick dotted line
is that with $\zeta_{\rm th} =1.2$ and $n=1.343$.  
The vertical thin solid line is the location of $M_{\rm
freeze}$ and the vertical thin dotted line is $M_{\rm RC} = 3.5
\times10^{13}$g. The PBHs with $M_{\rm RC}$ evaporate away at the
recombination epoch, $z\approx 1000$.  Because PBHs with masses smaller
than $M_{\rm freeze}$ have evaporated away before the redshift $z_{\rm
freeze}$, they provide no contribution to the $y$-distortion.  
It is shown in Fig. \ref{fig:yypermass.eps} that 
the largest contribution to the $y$-distortion is from PBHs with mass 
$M_{\rm freeze}$.  

\begin{figure}[htbp]
  \begin{center}
    \includegraphics[keepaspectratio=true,height=50mm]{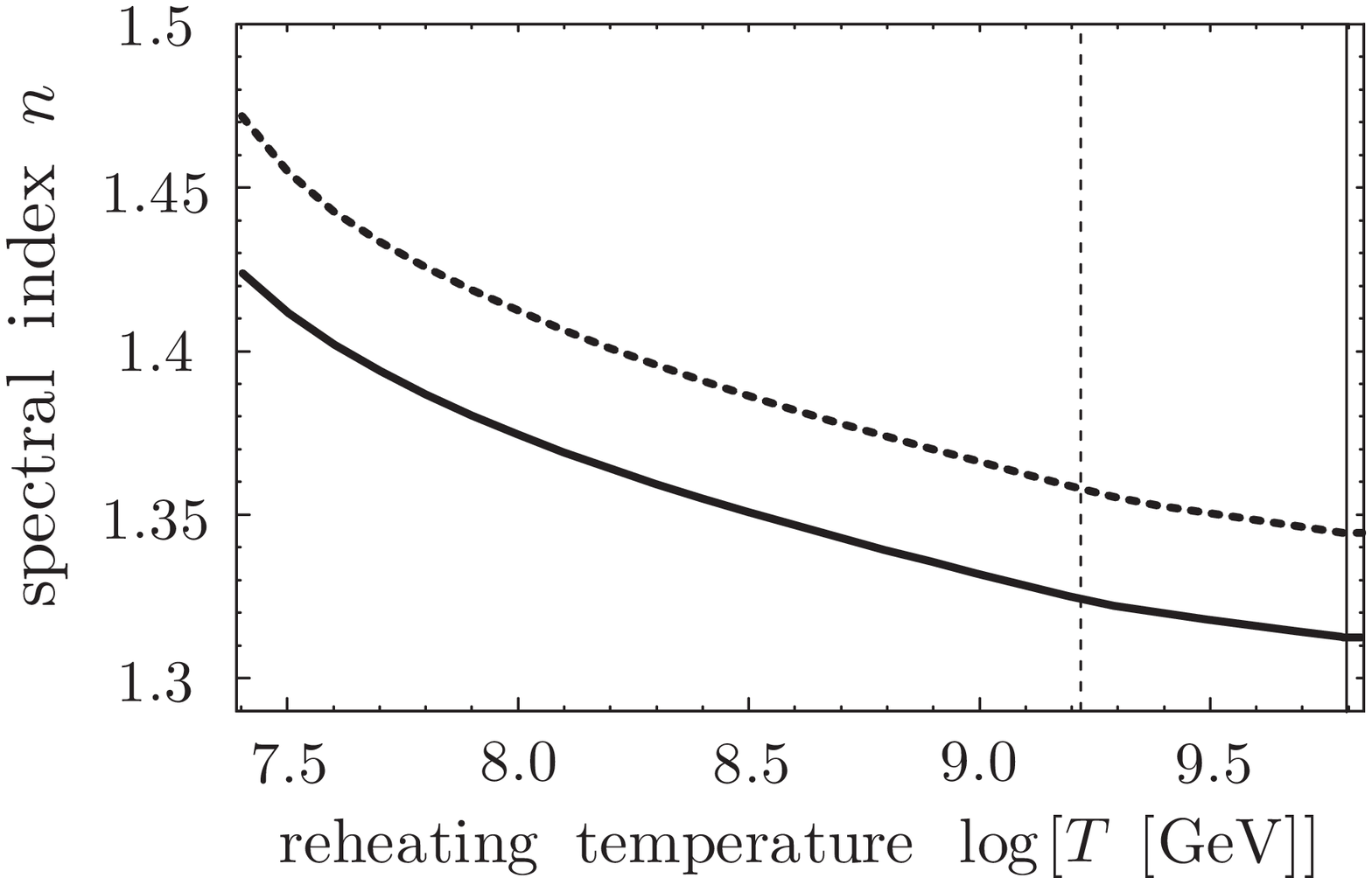}
  \end{center}
  \caption{The constraint on the spectral index as a function of the
   reheating temperature from the $y$-distortion. The thick solid
   line and the thick dotted line are the upper limits with the PBH
   critical threshold $\zeta_{\rm th} =0.7$ and $\zeta_{\rm th} =1.2$,
   respectively. The vertical thin dotted and thin solid lines
   represent the reheating temperatures whose horizon scales
   correspond to PBHs with masses $M_{\rm RC}$ and $M_{\rm freeze}$,
   respectively.  }
  \label{fig:yyreheat.eps}
\end{figure}

Next, we calculate the $y$-distortion in the case of the reheating
temperature $T_{\rm rh} \lesssim T_{\rm freeze}$. We plot constraints
on the power law index $n$ for a given reheating temperature in
Fig. \ref{fig:yyreheat.eps}. The thick solid and thick dotted lines are
the constraint of the spectral index for $\zeta_{\rm th} =0.7$ and
$\zeta_{\rm th} =1.2$, respectively. The vertical thin solid and thin
dotted lines represent $T_{\rm freeze}$ and $T_{\rm RC}\equiv 1.7
\times 10^{9}$GeV which corresponds to $M_{\rm RC}$, respectively.
When the reheating temperature is smaller than $T_{\rm freeze}$, the
larger spectral index is allowed because contributions on the
$y$-distortion from PBHs with masses between $M_{\rm BH,min}(T_{\rm
rh})$ and $M_{\rm freeze}$ are missing similar to the case of the
chemical potential.

\begin{figure}[htbp]
  \begin{center}
    \includegraphics[keepaspectratio=true,height=50mm]{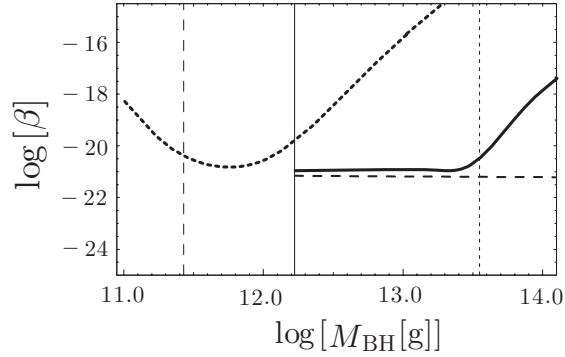}
  \end{center}
  \caption{The constraint on the PBH abundance from the $y$-distortion (thick solid line).
  The allowed region is under the thick solid line. 
  The vertical thin solid 
  and thin dotted lines represent the locations of $M_{\rm freeze}$ and
  $M_{\rm RC}$, respectively.  
  The thick dotted line is the constraint on the PBH abundance from the $\mu$-distortion.
  The thick dashed line is obtained by Eq.~(\ref{estimation}).  
  Here we adopt the COBE/FIRAS upper limit for $y$, and describe $z_{\rm e}$ and $z_{\rm BH}$ 
  in terms of $M_{\rm BH}$.
  }
  \label{fig:yybeta.eps}
\end{figure}

We calculate the upper bound of $\beta$ from the constraint on the
spectral index which is plotted in Fig. \ref{fig:yyreheat.eps}. We
plot the result as the solid line in Fig. \ref{fig:yybeta.eps}. As
a reference, we plot the constraint from the chemical potential as
the thick dotted line. These constraints, from $\mu$-distortion and
$y$-distortion, are complementary, as is shown in the figure. On the
mass scales $M_{\rm freeze} <M_{\rm BH}$, the $y$-distortion provides
a stringent constraint, while the $\mu$-distortion gives tighter constraint
on $M_{\rm freeze} >M_{\rm BH}$.  
We find that the upper bound of $\beta$ is $10^{-21}$ 
between $M_{\rm freeze}<M< M_{\rm RC}$.



Let us explain the above constraint on $\beta$ from the $y$-distortion
more intuitively. Most of the $y$-distortion is produced by PBHs with 
minimum mass at the redshift $z_{\rm freeze}$. Here we denote this minimum mass as $M_{\rm mini}$.  
If $T_{\rm rh} \gg T_{\rm freeze}$, $M_{\rm mini}=M_{\rm freeze}$ while 
$M_{\rm mini} \gtrsim M_{\rm freeze}$ for $T_{\rm rh} \lesssim T_{\rm freeze}$.  
We can approximate Eq. (\ref{y-dis}) as
\begin{equation}
y \approx  {Q (M_{\rm mini}) \over 4 \rho_r}   ,
\label{verify}
\end{equation}
where $Q (M_{\rm mini})$ is the total injected energy between $z_{\rm
freeze}$ and $z_{\rm e}$, which describes the epoch when the PBHs with
$M_{\rm mini}$ are evaporated away.
The maximum injected energy from the PBHs with $M_{\rm mini}$ is
$\rho_{\rm PBH } (M_{\rm mini})$. The temperature at formation of
PBHs with $M_{\rm BH}$ is $T_{\rm BH}\sim 3\times 10^9 {\rm GeV}
(M_{\rm BH}/10^{13})^{1/2}$ from Eq. (\ref{horizon-temperature}) so
that the formation redshift leads to $z_{\rm BH}\sim 10^{22} (M_{\rm
BH}/10^{13})^{1/2}$.  Since $\beta$ is defined as the fraction of
PBHs at the formation epoch, we must take into account the
evolution due to cosmic expansion to estimate the PBH fraction at
$z_e$. PBHs can be treated as a matter component while the universe
is dominated by radiation. Therefore the evolution factor between
$z_{\rm BH}$ and $z_{\rm e}$ can be written as $(z_{\rm BH}+1)/(z_{\rm e}+1)$.
Eventually we can relate the $y$-parameter to $\beta$ as 
\begin{equation}
y\lesssim  {\rho_{\rm PBH} (M_{\rm mini}) \over 4 \rho_r }
= {\Omega_{{\rm PBH}} \over 4}=
{z_{\rm BH}+1 \over z_{\rm e} +1} \beta ,
\label{estimation}
\end{equation}
where the inequality is introduced by
the fact that the maximum injected energy from PBHs 
provides only the upper bound for the $y$-distortion because 
PBHs can produce not only photons, but many kinds of particles, 
i.e., electrons, neutrinos, and so on. 
This relation is drawn in Fig. \ref{fig:yybeta.eps} as 
a thick dashed line. We find this intuitive approach is consistent with 
our previous constraint. 


\section{Constraint from the optical depth}

In this section,
we investigate the possible reionization of neutral hydrogens 
after recombination due to evaporation of PBHs.  

We adopt the cumulative number of ionizing photons 
which are produced by PBH evaporations 
as the criterion of the ionization.  
For simplicity, we ignore the cascade decays of high energy photons 
which can result in a larger number of photons. 
If we take this effect into account, the constraint 
on $\beta$ which we will obtain below, shall be much tighter.  
 
To achieve a volume-weighted ionization fraction of 99 percent, it is
found that 5 to 20 cumulative photons per hydrogen atom are needed \cite{h-a-m,
sokasian}. The particle emission rate of PBHs is given by
Eq. (\ref{emit-rate}) so that the total cumulative ionizing photon
number per hydrogen atom for given time $t$ is described as
\begin{equation}
{n_{\rm cum} \over n_{\rm H}}(t) = {1 \over n_{\rm H} }\int^{M_H(t)}
_{M_{\rm min}(t)} dM_{\rm BH} \int^{\infty}_{0} dE~n_{\rm BH}(M_{\rm
BH}) {dN_{\rm emit} \over dt dE}( M(M_{\rm BH},t)),
\label{number-injection}
\end{equation}
where $n_{\rm H}$ is the comoving hydrogen number density. Adopting
the upper bound of the power law index $n$ from the previous section
as $n \lesssim 1.3$, we find that the cumulative ionization photon
number is so negligible ($\sim 10^{-8}$ per hydrogen atom) that
reionization cannot be caused by ionizing photons injected by PBHs
unless the cascade decays of high energy photons take place very
efficiently such that one energetic photon decay into, say, $10^8$
ionizing photons.

\section{Conclusion}

In this paper, we investigated the possible influence of PBH evaporation on 
CMB. We set new and stringent constraints on the PBH mass fraction 
and the primordial power spectrum of density fluctuations from 
the observational upper bounds of $\mu$ and $y$-distortions obtained by 
COBE/FIRAS. For this analysis, we employed the peak theory, 
with a threshold described as $\zeta_{\rm th}$.  

It was shown that $\mu$ and $y$-distortions set limits on
different ranges of PBH masses. From the $\mu$-distortion, 
we can test the mass range between $ 2.7\times 10^{11} {\rm g}$ and $ 1.6
\times 10^{12}$g, which correspond to $M_{\rm DC}$ and $M_{\rm freeze}$, 
that is, the masses of PBHs 
which evaporate away at the decoupling epochs of the double Compton
and the thermalization, respectively.  
We obtain constraints on the power law spectral index 
as $n<1.304$ and $n<1.333$ for $\zeta_{\rm th}
=0.7$ and $\zeta_{\rm th} =1.2$, respectively. 
For the PBH abundance, we set the limit 
$\beta<10^{-21}$ for the mass range we considered here. 

On the other hand, from the $y$-distortion, 
we investigated the PBH mass range between 
$1.6 \times 10^{12} {\rm g}$ and $3.5\times 10^{13}$g, 
which correspond to $M_{\rm freeze}$ and $M_{\rm RC}$.
Here $M_{\rm RC}$ corresponds to the mass of PBHs which 
evaporate away at the recombination epoch.  
We obtained constraints on the power law spectral index 
as $n<1.312$ and $n<1.343$ for $\zeta_{\rm th}
=0.7$ and $\zeta_{\rm th} =1.2$, respectively. 
For the PBH abundance, we set the limit 
$\beta<10^{-21}$ for the mass range considered here.


It turns out that our constraints on the spectral index are looser than
those obtained from the observation of CMB temperature fluctuations
by WMAP satellite and other large scale observations such as galaxy
surveys of 2dF and SDSS, and surveys of Ly-$\alpha$ forest, which
imply a spectral index $n = 0.947 \pm 0.015$ \cite{wmap}. However, it
is suggested by many authors that general inflation models
produce not a simple power law spectrum but a spectrum with a running
spectral index which has many branches. Accordingly it may be no use 
to constrain the power law index from observations on various
scales. All we can do is to set constraints on the amplitude of each
scale associated with the observation. From this point of view, the
mass fraction of PBHs $\beta$ can provide unique information for the
fluctuation amplitude on very small scales, $M_{\rm DC} < M < M_{\rm
RC}$ which corresponds to the comoving scales between $3 \times
10^{-18}$Mpc and $ 4\times 10^{-17}$Mpc. Since these small scales
correspond to the Horizon scales right after the end of inflation, we
can say that the last stage of inflation can by revealed by PBHs.


Finally, we would like to mention that PBHs can be formed not only
by the primordial density perturbations but also by the
collisions of bubbles of the broken symmetry phase \cite{bubble} or by the
collapse of cosmic strings \cite{string}. Our constraint on $\beta$
is even applicable for such PBHs.  

\acknowledgements
NS is supported by a Grant-in-Aid
for Scientific Research from the Japanese Ministry of Education (No. 17540276).
NS thanks Institut d'astrophysique spatiale, 
Universit\'e Paris-Sud 11 for their kind hospitality.


\begin{thebibliography}{99}

\bibitem{wmap}
D. N. Spergel et al., Astrophys. J. S., {\bf 148}, 175 (2003).

\bibitem{2df}
W. J. Percival et al., Mon. Not. Roy. Astron. Soc. {\bf 327 }, 1297 (2001).

\bibitem{sdss}
M. Tegmark et al., Astrophys. J. {\bf 606}, 702 (2004).

\bibitem{spergel}
D. N. Spergel et al., Astrophys. J. S., {\bf 170}, 377 (2007).


\bibitem{pbh}
B. J. Carr and S. W. Hawking, Mon. Not. Roy. Astron. Soc. {\bf 168 }, 399 (1974).

\bibitem{hawkingrad}
S. W. Hawking, Nature, {\bf 248}, 30 (1974);\\
S. W. Hawking, Commun. Math. Phsy. {\bf 43}, 199 (1975).  




\bibitem{beta}
B. J. Carr. Astrophys. J. {\bf 201}, 1 (1975).



\bibitem{gamma}
D. Page and S. Hawking, Astrophys. J. {\bf 206}, 1 (1976);\\
J. H. MacGibbon and B. J. Carr, Astrophys. J. {\bf 371}, 447 (1991);\\
B. J. Carr and J. H. MacGibbon, Phys. Rep. {\bf 307}, 141 (1998);\\
E. V. Bugaev and K. V. Konishchev, Phys. Rev. D {\bf 66}, 084004 (2002).

\bibitem{nucle}
S. Miyama and K. Sato, Prog. Theor. Phys. {\bf 59}, 1012 (1978);\\
Ya. B. Zel'dovich, A. A. Starobinsky, M. Iu. Khlopov, and V. M. Chechetkin,\\ 
Pis'ma Astron. Zh. {\bf 3}, 208 (1977) [Sov. Astron. Lett. {\bf 3}, 110 (1977)];\\
D. Lindley, Mon. Not. R. Astron. Soc. {\bf 193}, 593 (1980);\\
T. Rothman and R. Matzner, Astrophys. Space Sci. {\bf 75}, 229 (1981);\\
K. Kohri and J. Yokoyama, 
Phys.\ Rev.\  D {\bf 61}, 023501 (1999). 


\bibitem{abundance}
B. J. Carr and J. H. MacGibbon, Phys. Rep. {\bf 307}, 141 (1998);\\
B. J. Carr, J. H. Gilbert and J. E. Lidsey, Phys. Rev. D {\bf 50} 4853 (1994);\\
B. J. Carr and J. E. Lidsey, Phys. Rev. D {\bf 48} 543 (1993);\\
A. M. Green and A. R. Liddle, Phys.Rev. D {\bf 56} 6166 (1997);\\
A. M. Green and A. R. Liddle, Phys.Rept. {\bf 307} 125 (1998).


\bibitem{distort}
P. D. Naselskii, Pisma Astron. Zh. {\bf 4}, 387 (1978) [Sov. Astron. Lett. {\bf 4}, 209 (1978)].

\bibitem{naselskii-shevelev}
P. D. Naselskii and Y. G. Shevelev, Astrophysics {\bf 14}, 386 (1978).

\bibitem{ricotti}
M. Ricotti, J. P. Ostriker, and K. J. Mack, arXiv0709.0524 (2007).


\bibitem{firas}
J. C. Mather et.al., Astrophys. J. {\bf 420} 439 (1994).

\bibitem{n-j}
J.C. Niemeyer and K. Jedamzik, Phys. Rev. Lett. {\bf 80}, 5481 (1998).

\bibitem{s-s}
M. Shibata and M. Sasaki, Phys. Rev. D {\bf 60}, 084002 (1999). 

\bibitem{g-etal}
A. M. Green, A. R. Liddle, K. A. Malik and M. Sasaki, Phys. Rev. D {\bf 70}, 041502(R) (2004).




\bibitem{invariant}
D. Wands, K. A. Malik, D. H. Lyth, and A. R. Liddle,
Phys. Rev. D {\bf 62}, 043527 (2000).

\bibitem{lightman}
A. P. Lightman, Astrophys. J. {\bf 244}, 392 (1981).

\bibitem{h-s}
W. Hu and J. Silk, Phys. Rev. D {\bf 48}, 485 (1993).

\bibitem{mac-w}
J. H. MacGibbon and B. R. Webber, Phys. Rev. D {\bf 41}, 3052 (1990).

\bibitem{macgibbon}
J. H. MacGibbon, Phys. Rev. D {\bf 44}, 376 (1991).

\bibitem{s-z}
Y. B. Zeldovich and R. A. Sunyaev, Ap. Space Sci., {\bf 4}, 301 (1969).

\bibitem{f-k}
M. Fukugita and M. Kawasaki, Astrophys. J. {\bf 353}, 384 (1990).

\bibitem{h-a-m}
Z. Haiman, T. Abel and P. Madau, Astrophys. J. {\bf 551}, 599 (2001).

\bibitem{sokasian}
A. Sokasian, T. Abel, L. Hernquist and V. Springel, Mon. Not. Roy. Astron. Soc. {\bf 344}, 607, (2003);\\
A. Sokasian, N. Yoshida, T. Abel, L. Hernquist and V. Springel, Mon. Not. Roy. Astron. Soc. {\bf 350}, 47, (2004).



\bibitem{bubble}
S. W. Hawking, I. G. Moss, J. M. Stewart, Phys. Rev. D {\bf 26} 2681 (1982);\\
I. G. Moss, Phys. Rev. D {\bf 50} 676 (1994);\\
R. V. Konoplich, S. G. Rubin, A. S. Sakharov and M. Yu. Khlopov, Phys. Atom. Nucl. {\bf 62} 1593 (1999).

\bibitem{string}
S. W. Hawking, Phys. Lett. B 231, 237 (1989);\\
A. Polnarev and R. Zembowicz, Phys. Rev. D {\bf 43}, 1106 (1991);\\
J. Garriga and M. Sakellariadou, Phys. Rev. D {\bf 48} 2502 (1993);\\ 
R. R. Caldwell and P. Casper, Phys. Rev. D {\bf 53}, 3002 (1996);\\
J. H. MacGibbon, R. H. Brandenberger and U. F. Wichoski, Phys. Rev. D {\bf 57} 2158 (1998).


\end{thebibliography}
\end{document}